\documentclass[12pt,a4paper,twoside]{article}

\usepackage{amssymb}
\usepackage{latexsym}
\usepackage{theorem}
\usepackage{amsfonts,amsmath}

\newcommand{\R}{{\mathord{\mathbb R}}}

\newcommand{\C}{{\mathord{\mathbb C}}}

\newcommand{\mH}{{\mathcal H}}

\newcommand{\mL}{{\mathcal L}}

\newcommand{\mO}{{\mathcal O}}

\newcommand{\tr}{{\rm tr}}

\newcommand{\CAR}{{\mathfrak F}}

\begin{document}


\pagestyle{myheadings}

\markboth{W.H. Aschbacher, H. Spohn}{A short remark on the strict positivity of the entropy production}

\title{A Remark on the Strict Positivity of the Entropy Production}

\author{Walter H. Aschbacher, Herbert Spohn\\ \\
Technische Universit\"at M\"unchen \\ 
Zentrum Mathematik\\
85747 Garching, Germany\\
email: {\tt aschbacher@ma.tum.de,spohn@ma.tum.de}}

\maketitle


\begin{abstract}
We establish an algebraic criterion which ensures the strict positivity of the entropy production in quantum models consisting of a small system coupled to thermal reservoirs at different temperatures. 
\end{abstract}

\section{Introduction}

When a small quantum system is coupled to two ideal infinitely extended heat reservoirs at different temperatures, physically one would expect to have a {\it non-zero steady state energy flux} directed from the hot to the cold reservoir. To establish such a property on the basis of a microscopic Hamiltonian, including the thermal reservoirs, is not so obvious, since, in principle, the energy flux could vanish because of two obstructions:\medskip\\
$(i)$ The coupling to the reservoirs may be ineffective.\\
$(ii)$ Inside the small system there could be an unsurmountable energy flux barrier. \medskip\\
The aim of our note is to establish a manageable criterion which ensures a strictly positive entropy production, in other words a nonvanishing energy flux. 

To be more precise, we consider a small system with a finite dimensional Hilbert space $\mH$ coupled to two identical reservoirs consisting of ideal Fermi gases, with fermionic Fock space $\CAR$.  The total Hamiltonian on $\mH\otimes\CAR\otimes\CAR$ is of the form
\begin{eqnarray}
\label{Hlambda}
H(\lambda)&=&H\otimes 1\otimes 1+1\otimes H_{f}\otimes 1+1\otimes 1\otimes H_f\nonumber\\
&+&\lambda \, \left(Q_L\otimes \varphi(\alpha_L)\otimes 1+Q_R\otimes 1\otimes  \varphi(\alpha_R)\right),
\end{eqnarray}  
where $H$ is the Hamiltonian of the small system, $Q_L$, $Q_R$ are the self-adjoint coupling operators acting on $\mH$, and $H_f$ is the free field Hamiltonian of the reservoir acting on $\CAR$. The coupling to the reservoirs is linear in the field operators and smeared over suitable test functions $\alpha_L, \alpha_R$. Since the thermal reservoirs are infinitely extended, one has to take suitable limits. A more direct approach is to construct,  in the sense of \cite{R01},
the non-equilibrium steady state (NESS) within the mathematical framework of algebraic quantum statistical mechanics. We refer to  \cite{AJPP}, \cite{JP1}  for a more detailed discussion. In this framework, each part of the total system, i.e. the small system and the reservoirs, is described by a $C^\ast$-dynamical system, i.e. a  $C^\ast$-algebra of observables and a group of time evolution automorphisms on the observables. If we assume the initial conditions to be such that both the small  system and the reservoirs are in thermal equilibrium (for example infinite temperature for the small system as in \cite{JP2} and $T_L$, $T_R$ for the reservoirs), then the GNS-representation (w.r.t. this initial condition) allows us to treat the $C^\ast$-dynamical system within Hilbert space
formalism. In this Hilbert space the time evolution is implemented through a unitary group generated by a self-adjoint operator, the so called standard Liouvillian, constructed by means of the Tomita-Takesaki modular theory of von Neumann algebras. If the system is close to thermal equilibrium the spectral theory of the standard Liouvillian encodes the ergodic properties of the system. In contrast, if the system is far from equilibrium, the standard Liouvillian does not provide readily accessible infinite volume information. However, using again the Tomita-Takesaki modular theory, the so called $C$-Liouvillian is introduced in \cite{JP2}, a non-self-adjoint operator which generates a non-unitary group implementing  the time evolution on the Hilbert space. Complex deformation techniques allow then for a relation between NESS and zero-resonance eigenvectors of this $C$-Liouvillian. 

In the generality posed, to establish the existence of  a non-zero steady state energy flux seems rather intractable. However, if the coupling constant $\lambda$ is small but finite, say  $|\lambda|\le \lambda_0$ with sufficiently small $\lambda_0>0$, there has been considerable progress recently \cite{JP2}. In particular, if the system is effectively coupled, see $(i)$ below, and under suitable assumptions on the regularity of the form factors $\alpha_L$, $\alpha_R$ (because of the use of the complex deformation techniques), it can be proven that, for $0<|\lambda|\le\lambda_0$, the NESS is unique and has a perturbation expansion in $\lambda$.  Hence, the steady state energy flux oriented from the left reservoir into the small system reads 
\begin{eqnarray}
\label{flux}
j(\lambda)=\lambda^2 \sigma_0+\mO(\lambda^3).
\end{eqnarray}
Moreover, $\sigma_0$ is computable in terms of the Davies weak coupling generator on the Hilbert space $\mH$ of the small system. Therefore, the issue whether $j(\lambda)\neq 0$ for $0<|\lambda|\le \lambda_0$ is reduced to a much simpler question, namely whether for the dissipative quantum dynamics of the small system the steady state energy flux $\sigma_0$ does not vanish. 

In this note, our contribution is to provide a simple algebraic criterion which ensures $\sigma_0\neq 0$. Physically, the criterion expresses that the system will thermalize at $\beta_L$ when coupled {\it only} to the left reservoir, and alike for $\beta_R$. When coupled to both reservoirs at different temperatures it follows that $\sigma_0\neq 0$. This condition is expected to be sufficient only and we will confirm so for an explicit example in the last section.

Due to the conservation of energy, the entropy production $\epsilon(\lambda)$ in the NESS can be written as 
\begin{equation}
\label{entropy}
\epsilon(\lambda)=\lambda^2 (\beta_R\!-\!\beta_L) \,\sigma_0+{\mathcal O}(\lambda^3).
\end{equation}
Thus, equivalently, we state a sufficient condition for the entropy production to be strictly positive provided $0<|\lambda|\le \lambda_0$.

For classical Hamiltonian models the same problem has been investigated in \cite{EPR99}. There, a rather specific form of the smearing functions $\alpha_L$, $\alpha_R$ is required. But given this restriction, the result in \cite{EPR99} is more general than the one proved here. In particular, there is no condition of small coupling, thus no recourse to a weak coupling effective dynamics.

\section{The Davies generator}

In the following, we briefly explain the quantum dynamical semigroup for the effective dynamics of the small system at weak coupling. For a more detailed description see \cite{D74,LS}.

Let us define  the Fourier transform of the time correlation functions of the reservoir part of the interactions ($r$ will always stand for the left or the right reservoir, $r=L,R$),
\begin{equation}
\label{hr}
h_r(E)=\int_{-\infty}^\infty dt\,\,e^{-iEt}\,\omega_r(\varphi(\alpha_r)\,\tau_r^t\varphi(\alpha_r)).
\end{equation}
Here $\omega_r$ denotes the thermal equilibrium state of reservoir $r$ w.r.t. to the time evolution $\tau_r^t$ of the ideal Fermi gas at inverse temperature $\beta_r$ and $\varphi(\alpha_r)$ stems from \eqref{Hlambda}.  Note that, since $\omega_r$ is a $(\tau_r,\beta_r)$-KMS state, $h_r(E)$ has the property (\cite[(III.16)]{LS}) \begin{equation}
\label{kms}
h_r(-E)=e^{-E\beta_r } \,h_r(E)\ge 0.
\end{equation}
A natural condition is to assume that the reservoirs induce transitions between any two energy levels of the small system. Denoting by $\sigma(A)$ the spectrum of the operator $A$, this leads to\medskip\\
{\bf Assumption} $\mathrm{(E_r)}$ \,({\it Effective  coupling})
\begin{equation*}
h_r(E)>0\,\,\,\mbox{for all} \,\,\, E\in\sigma([H,\cdot]).\medskip
\end{equation*}

To construct the Davies generator, let $E_n\in\sigma(H)$ and $P_n$ be the corresponding spectral projection. Then $H=\sum_nE_nP_n$. Moreover, let $E\in \sigma([H,\cdot])$ with $[H,\cdot]$ the Liouvillean for $H$. With the help of the definition
\begin{equation}
\label{Vomega}
Q_r(E)=\sum_{E_m-E_n=E}P_nQ_rP_m
\end{equation}
we can write the Davies generator $K_r$ in the form 
\begin{eqnarray}
\label{Kr}
K_r\rho&=&\hspace{-0.3cm}\sum_{E\in\sigma([H,\cdot])}\hspace{-0.2cm}-is_r(E)\,[Q_r(E)^\ast Q_r(E),\rho\,]\nonumber\\
&&\hspace{1cm}+h_r(E)\left([Q_r(E)\rho,Q_r(E)^\ast]+[Q_r(E),\rho \,Q_r(E)^\ast]\right).
\end{eqnarray}
Here, $K_r\in\mL(\mL^1(\mH))$, i.e. $K_r$ is a bounded operator on the trace class operators $\mL^1(\mH)$ on $\mH$, $h_r$ is defined in \eqref{hr}, and $s_r(E)= 1/2\pi\, pv \int_{-\infty}^\infty dE'\,\,h_r(E')/(E'-E)$ denotes the Hilbert transform of $h_r$. If the small system is coupled to both reservoirs the generator $K$ in the weak coupling limit is the sum 
\begin{eqnarray}
\label{K}
K=K_L+K_R.
\end{eqnarray}
A steady state $\rho_0$ of the Davies generator $K$ is dermined by
\begin{eqnarray}
\label{Krho0}
K\rho_0=0.
\end{eqnarray}
Theorem 3 in \cite{LS} asserts that, if $\mathrm{(E_L)}$ and $\mathrm{(E_R)}$ hold, and if the commutant $\{H,Q_L,Q_R\}'=\C 1$, then \eqref{Krho0} has a unique solution \,(we denote by $X'=\{y\in \mL(\mH)\,|\,[y,x]=0\,\,\mbox{for all}\,\, x\in X\}$ the commutant of the subset $X\subseteq \mL(\mH)$). In fact, under this condition, and provided $0<|\lambda|\le \lambda_0$, it is proved that the full microscopic model converges to a unique NESS as $t\to\infty$.  

In the weak coupling approximation the change of energy is
\begin{eqnarray*}
\frac{d}{dt}\,\tr(H\rho(t))=\tr(HK\rho(t)),
\end{eqnarray*}
and, thus,  the steady state flux $\sigma_0$ from \eqref{flux}, \eqref{entropy} should be $\tr(HK_L\rho_0)$. Indeed, from \cite{JP1},\cite{JP2},
\begin{eqnarray*}
\sigma_0=\tr(HK_L\rho_0).
\end{eqnarray*}
The issue is to have a condition ensuring $\sigma_0>0$ in case $\beta_R-\beta_L>0$.

\section{Strict positivity of the entropy production}

The thermalization at coupling only to one reservoir is guaranteed by\medskip\\
{\bf Assumption} $\mathrm{(C_r)}$ \,({\it Triviality of commutants})
\begin{equation*}
\{H, Q_r\}'=\C\,1.\medskip
\end{equation*}
We now state our claim.\medskip\\
{\bf Theorem}\,\,(Strict positivity of entropy production). {\it
Let the small system be well coupled in the sense of $\mathrm{(E_L)}$, $\mathrm{(E_R)}$, let $\beta_L\neq\beta_R$, and let $\mathrm{(C_L)}$ and $\mathrm{(C_R)}$ hold. 
Then,  for sufficiently small $\lambda$, the entropy production satisfies}
\begin{eqnarray*}
\epsilon(\lambda)>0.\medskip
\end{eqnarray*}
{\it Remark.} The conditions are not necessary as can be seen from the example {\it (i)} in the last 
section.\medskip\\
{\bf Corollary}\,\,(Non-zero steady state energy flux)
{\it Under the same conditions, if $\beta_R>\beta_L$, the steady state energy flux satisfies }
\begin{eqnarray*}
j(\lambda)>0.
\end{eqnarray*}
{\bf Proof}\,\,We apply Theorem 3 from \cite{LS} for a single reservoir (see also \cite{S1},\cite{S2}).  This theorem proceeds on the assumption that  the small system coupled to reservoir $r$  has a Davies generator $K_r$ of the form \eqref{Kr}. It then asserts that, if $\mathrm{(E_r)}$ holds, $\mathrm{(C_r)}$ implies $\dim\ker K_r=1$. Since the thermal Gibbs state $\rho_r=e^{-\beta_r H}/\tr(e^{-\beta_r H})$ is stationary (cf. \cite[(III.22)]{LS}), the only density matrix $\rho$ on $\mH$ which solves  $K_r\rho=0$ is $\rho=\rho_r$.

Next, we want to take advantage of part $(i)$ of Theorem 2 in \cite{LS} about the relation of the kernel of the Davies generator $K_r$ and the vanishing of   the entropy production $\sigma_r(\rho)$. In \cite[(V.6)]{LS} the  entropy production $\sigma_r(\rho)$ on the state $\rho$ of the small system coupled to reservoir $r$ is defined as the change of the relative entropy w.r.t. the thermal Gibbs state along the trajectory of the time evolved density matrix $\rho$. Furthermore, from \cite[(V.29)]{LS} we know that   $\sigma_r(\rho)$ has the form
\begin{eqnarray*}
\sigma_r(\rho)=-\beta_r \,\tr(H K_r\rho)-\tr(\log\!\rho\,K_r\rho).
\end{eqnarray*}
Now, part $(i)$ of Theorem 2 in  \cite{LS} asserts that $\sigma_r(\rho)$ is nonnegative, and, under $\mathrm{(E_r)}$, the  entropy production $\sigma_r(\rho)$ vanishes only for $\rho=\rho_r$.
Furthermore, due to \eqref{K}, the total entropy production $\sigma(\rho)$, i.e. the entropy production if the small system is coupled to both reservoirs, can be written as
\begin{eqnarray*}
\sigma(\rho)=\sigma_L(\rho)+\sigma_R(\rho).
\end{eqnarray*}
Since the temperatures of the two reservoirs are different, 
$\rho_L\neq\rho_R$ . 
By $\mathrm{(C_L)}$ and $\mathrm{(C_R)}$, $\{H,Q_L,Q_R\}=\C 1$  and we can again apply Theorem 3 from \cite{LS} which implies that $\dim \,\ker K=1$,  $K\rho_0=0$.
Therefore, the total entropy production $\sigma(\rho_0)$ is strictly positive,
\begin{eqnarray*}
\sigma(\rho_0)>0.
\end{eqnarray*}
Furthermore, it is of the form $\sigma(\rho_0)=-\beta_L \,\tr(H K_L\rho_0)-\beta_R \,\tr(H K_R\rho_0)$, since,  by stationarity of $\rho_0$, the contribution  $\tr(\log\!\rho_0\,K\rho_0)$ vanishes. Finally, $\sigma_0$ in \eqref{flux}, \eqref{entropy} is given by $\sigma(\rho_0)=(\beta_R-\beta_L)\sigma_0$.

\hfill $\Box$


\section{Applications}
As an illustration, we discuss two simple  examples for the small system.\medskip\\
{\bf A single spin}\smallskip\\
Let the small system consist of a single spin $1/2$ with Hilbert space $\mH=\C^2$ and Pauli matrices $\sigma_1, \sigma_2, \sigma_3$. Then, the condition $\mathrm{(C_r)}$ is equivalent to $[H,Q_r]\neq 0$. Our theorem hence implies e.g. that  the single spin with $H=\sigma_3$ and $Q_L=Q_R=\sigma_1$ has strictly positive entropy production which is also established in \cite{JP1}, \cite{JP2}.\medskip\\
{\bf Two $XY$ coupled spins}\smallskip\\
We consider the two-spin Hamiltonian of $XY$ type ($\gamma_1,\gamma_2\in\R$),
\begin{eqnarray*}
H=\frac{1}{2}\left(\sigma_3\otimes 1+1\otimes \sigma_3+\gamma_1\,\sigma_1\otimes\sigma_1+\gamma_2\,\sigma_2\otimes\sigma_2\right).
\end{eqnarray*}
The coupling operators $Q_L$, $Q_R$ are chosen to be of the form
\begin{eqnarray*}
Q_L=\sigma_1\otimes 1,\quad Q_R=1\otimes \sigma_1.
\end{eqnarray*}
We want to discuss three choices for the parameters $\gamma_1,\gamma_2$. \medskip\\
{\it (i)} \,$\gamma_1\!=\!\gamma_2\!=\!1$.
In this case, 
$$
1\otimes\sigma_1+\sigma_1\otimes\sigma_3\in \{H,Q_L\}',\quad \sigma_1\otimes 1+\sigma_3\otimes\sigma_1\in \{H,Q_R\}'.
$$
Hence, the assumptions  $\mathrm{(C_L)}$ and $\mathrm{(C_R)}$ do not hold. By direct computation we find that \,$\dim\ker K_r=2$, imposing only $\eqref{kms}$.
Nevertheless the commutant $\{H,Q_L,Q_R\}'$ is trivial and the Davies generator $K$ has a unique stationary state $\rho_0$ which turns out to be of the form 
\begin{eqnarray*}
\rho_0=\frac{1}{4}\Big( 1\otimes 1-\frac{\sinh(\beta_L\!+\!\beta_R)}{2\cosh\beta_L\cosh\beta_R}\, H\Big),
\end{eqnarray*}
independent of the choice for $h_r$ and $s_r$. Calculating its entropy production we find
$$
\sigma(\rho_0)=(\beta_R\!-\!\beta_L)\frac{\sinh (\beta_R\!-\!\beta_L)}{\cosh\beta_L\cosh\beta_R}>0,
$$
which is stricly positive. This example shows that, in general, the conditions of our theorem are not necessary. 
\medskip\\
{\it (ii) \,anisotropic $XY$ coupling}. If we set $\gamma_1=1+\gamma$ and $\gamma_2=1-\gamma$, then $\gamma_1=\gamma_2=1$ corresponds to $\gamma=0$. For $\gamma\neq 0$, the anisotropic $XY$ coupling, the commutants are trivial, i.e. $\mathrm{(C_L)}$ and  $\mathrm{(C_R)}$ hold. Hence, our theorem is applicable.
\medskip\\
{\it Remark}.
The existence of a non-zero energy flux through the infinite $XY$ chain can be proved using scattering theory on the one particle Hilbert space of the free Fermion system arising from the $XY$ chain under a Jordan-Wigner transformation, cf. \cite{AP03}. \medskip\\
{\it (iii) \,$XY$ chain cut apart, \,$\gamma_1\!=\!\gamma_2\!=\!0$.}
The right and left system are uncoupled and, clearly, there is no heat flux at any $\lambda$. To see how our theorem fails, one notes that $\mathrm{(C_r)}$ does not hold. $\{H,Q_L,Q_R\}'=\C 1$ is in force, and there is a unique stationary state $\rho_0$ of $K$ given by $(2\cosh (\beta_L/2)2\cosh (\beta_R/2))^{-1}e^{-(\beta_L/2)\sigma_3}\otimes e^{-(\beta_R/2)\sigma_3}$.




\begin{thebibliography}{99}

\bibitem{AJPP}Aschbacher, W.H., Jak\v si\'c, V., Pautrat, Y., Pillet, C.-A.: {\it Topics in non-equilibrium quantum statistical mechanics}, Springer Lecture Notes in Mathematics, to appear, mp\_arc 05-69

\bibitem{AP03}Aschbacher, W.H., Pillet, C.-A.: {\it Non-equilibrium steady states of the XY chain}, J. Stat. Phys. {\bf 112} (2003), 1153

\bibitem{D74}Davies, E.B.: {\it Markovian master equations}, Comm. Math. Phys. {\bf 39} (1974), 91

\bibitem{EPR99}Eckmann, J.-P., Pillet, C.-A., Rey-Bellet, L.: {\it Entropy production in non-linear, thermally driven Hamiltronian systems}, J. Stat. Phys. {\bf 95} (1999), 305

\bibitem{JP1}Jak\v si\'c, V., Pillet, C.-A.: {\it Mathematical theory of non-equilibrium quantum statistical mechanics}, J. Stat. Phys. {\bf 108} (2002), 787

\bibitem{JP2}Jak\v si\'c, V., Pillet, C.-A.: {\it Non-equilibrium steady states of finite quantum systems coupled to thermal reservoirs}, Comm. Math. Phys. {\bf 226} (2002), 131

\bibitem{R01}Ruelle, D.: {\it Entropy production in quantum spin systems}, Comm. Math. Phys. {\bf 224} (2001), 3

\bibitem{S1}Spohn, H.: {\it Approach to equilibrium for completely positive dynamical semigroups of n-level systems}, Rep. Math. Phys. {\bf 10} (1976), 189

\bibitem{S2}Spohn, H.: {\it An algebraic condition for the approach to equilibrium of an open n-level system}, Lett.  Math. Phys. {\bf 2} (1977), 33

\bibitem{LS}Spohn, H., Lebowitz, J.L.: {\it Irreversible thermodynamics for quantum systems weakly coupled to thermal reservoirs}, Adv. Chem. Phys. {\bf 38} (1978), 109




\end{thebibliography}
\end{document}